# A Correction to the Immirizi Parameter of *SU*(2) Spin Networks

M. Sadiq[1], Department of Physics, Universsity of Tabuk, P.O. Box 2072,Tabuk 71451, Saudi Arabia

**Abstract**

The elegant predictions of loop quantum gravity are obscured by the free Immirizi parameter ($\gamma$). Dreyer (2003), considering the asymptotic quasinormal modes spectrum of a black hole, proposed that $\gamma$ may be fixed by letting the *j*=1 transitions of spin networks as the dominant processes contributing to the black hole area, as opposed to the expected *j*=1/2 transitions. This suggested that the gauge group of the theory might be *SO*(3) rather than *SU*(2). Corichi (2003), maintaining *SU*(2) as the underlying gauge group, and invoking the principle of local fermion-number conservation, reported the same value of $\gamma$ for *j*=1 processes as obtained by Dreyer. In this note, preserving the *SU*(2) structure of the theory, and considering *j*=1 transitions as the dominant processes, we point out that the value of $\gamma$ is in fact twice the value reported by these authors. We arrive at this result by assuming the asymptotic quasinormal modes themselves as dynamical systems obeying *SU*(2) symmetry.

## 1. Introduction

Loop quantum gravity (LQG) has produced results from first principle that geometry is discrete. LQG uses spin networks as a basis for its Hilbert space. Spin networks are graphs with edges that carry labels as $j = 0, 1/2, 1, 3/2,\ldots$, that is, the representations of *SU*(2) group that serves as the gauge group of the theory. In LQG the area of a given region of space has a discrete spectrum in such a way that if a surface is punctured by an edge of the spin network carrying a label *j*, the surface acquires an area element [1,2,3]

$$A_j = 8\pi l_P^2 \gamma \sqrt{j(j+1)}. \tag{1}$$

Here $l_P^2$ is the Planck area and $\gamma$ is the free undetermined Immirizi parameter [4] in the theory that remains obscured. The remarkable predictions of the theory are ambiguous up to this all-time present unfixed parameter. Nevertheless, indirect tools are used to fix the value of $\gamma$.

In [5,6], it was shown that $\gamma$ can be fixed by the requirement that the quantum gravity results reproduce the Bakenstein-Hawking entropy [7,8]

$$S = \frac{A}{4l_P^2}. \tag{2}$$

A systematic approach to quantum black hole entropy was used to fix the value of $\gamma$ as

$$\gamma = \frac{\ln(2j_{min}+1)}{\sqrt{j_{min}(j_{min}+1)}}, \qquad (3)$$

where $j_{min}$ is the minimum (semi-integer) label for the representations of $SU(2)$, responsible for the black hole entropy. Taking into consideration that statistically the most important contribution should come from $j_{min} = 1/2$, $\gamma$ is fixed as

$$\gamma = \frac{\ln 2}{\pi\sqrt{3}}. \qquad (4)$$

---
[1] e-mail: ms.khan@ut.edu.sa

Dreyer [9], following a clue uncovered by Hod [10], fixed the value of $\gamma$ in an independent way by using a semi-classical argument based on the quasinormal mode (QNM) spectrum of a Schwarzschild black hole [11,12]. Dreyer's approach was based on Hod's conjecture that the real part of the highly damped (QNM) frequencies $\omega_{QNM}$ asymptotically tends to a fixed quantity

$$\omega_{QNM} = \frac{\ln 3}{8\pi M}. \qquad (5)$$

This conjecture was proved analytically by Motl [13]. The argument used by Hod and also by Dreyer goes as follows: If we assume that the Bohr's correspondence principle is applicable to black holes, the radiation or absorption of such an asymptotic frequency of the quasinormal modes should be consistent with the variation in mass $\Delta M$ of the black hole, i.e.

$$\Delta M = \hbar \omega_{QNM} = \frac{\hbar \ln 3}{8\pi M}. \qquad (6)$$

Since the area $A$ of the event horizon and the mass $M$ of a Schwarzschild black hole are related by

$$A = 16\pi M^2, \qquad (7)$$

the variation in mass causes change $\Delta A$ in the quantized area of the event horizon. With the help of (5) and (6), equation (7) readily gives

$$\Delta A = 4 l_P^2 \ln 3. \qquad (8)$$

Dreyer considered that the most natural candidate for a transition of the quantum black hole, as described above, is the appearance or disappearance of a puncture with spin $j_{min}$. The area of the black hole would then change by an amount given by equation (1), where $j = j_{min}$, i.e.

$$\Delta A = A_{j_{min}} = 8\pi\gamma l_P^2 \sqrt{j_{min}(j_{min}+1)}. \qquad (9)$$

Comparison of (8) and (9) then yields the value of the Immirizi parameter as

$$\gamma = \frac{\ln(2j_{min}+1)}{2\pi\sqrt{j_{min}(j_{min}+1)}}. \qquad (10)$$

This value also sets the result for the black hole entropy as

$$S = \frac{A}{4l_P^2} \frac{\ln(2j_{min}+1)}{\ln 3}. \qquad (11)$$

In order to comply with the Bakenstein-Hawking entropy, Dreyer was forced to fix $j_{min} = 1$, and consequently, the Immirizi parameter as

$$\gamma = \frac{\ln 3}{2\pi\sqrt{2}} \quad (12)$$

At this point, in the absence of any other explanation why $j_{min}=1/2$ does not comply with the correct entropy formula, Dreyer proposed that the true gauge group of the theory might be $SO(3)$ rather than $SU(2)$.

There have been various attempts to formulate a convincing explanation about why $j=1$ processes contribute dominantly to the black hole entropy. Corichi [14] has argued that one should maintain $SU(2)$ as the gauge group of the theory if fermions are to be included in the theory. However, the $j_{min}=1/2$ processes has to be highly suppressed. Corichi's argument goes as follows. Losing a $j=1/2$ representation would mean the edge becomes open in the bulk. In order to keep the local gauge invariance intact one has to attach a fermion to the open end; and this is not allowed if the fermion number is to be conserved locally. On the other hand, if an edge carries $j = 1$, one could attach a fermion-antifermion pair to the open end of the detached edge, preserving the local gauge invariance. Thus, even though punctures with $j=1/2$ edges are allowed kinemetically, the dominant contribution comes from processes for which the minimum allowed value of $j$ is 1. Swain [15] proposed a generalized version of Pauli's exclusion principle applied to spin networks, which stated that "no more than two punctures of $j=1/2$, each with differing $m$ values, may puncture a given surface". In this perspective, even though $j=1/2$ punctures are not forbidden, the dominant contribution appears to be coming from $j=1$ punctures. Astonishingly, exact agreement with correct entropy formula was achieved for $j=1/2$ as the dominant contributing processes by invoking supersymmetric extension of spin networks [16].

In the present note, working in the framework of $SU(2)$ spin networks, we propose that the dynamics of an asymptotic QNM frequency be described by the Hamiltonian of a two-dimensional isotropic oscillator. It looks a natural choice because the 2D oscillator possesses the same group structure and symbols as those of the edges in loop quantum gravity. This picture clearly explains how the conversion of the quanta of geometry to matter quanta, and vice versa, at the horizon takes place in a consistent way. Further, as we will see, for $SU(2)$ as the gauge invariance, and $j=1$ as the dominant processes, the value of the Immirizi parameter explicitly turns out to be twice the value reported by Dreyer and Corichi based on $SO(3)$ and $SU(2)$, respectively.

For relevance, we start in section 2 with a brief review of the Schwinger's scheme [17] of realizing a 2D isotropic oscillator as $SU(2)$ system. In section 3 we work out the value of $\gamma$ for $j=1$ transitions in the $SU(2)$ framework. The conclusions are presented in section 4.

## 2. 2D Isotropic Oscillator and $SU(2)$ Symmetry

In this section we briefly review Schwinger's method of establishing the relationship between a two-dimensional isotropic oscillator and the $SU(2)$ symmetry. Consider a 2D isotropic oscillator described by the Hamiltonian (assuming $\hbar = 1$ in this section)

$$\hat{H} = \left(a_1^\dagger a_1 + a_2^\dagger a_2 + 1\right)\omega, \tag{13}$$

where $\omega$ is the frequency of the oscillator and the algebra

$$\left[a_i, a_j^\dagger\right] = \delta_{ij}, \quad \left[a_i, a_j\right] = \left[a_i^\dagger, a_j^\dagger\right] = 0, \quad i, j = 1, 2. \tag{14}$$

holds. The Hilbert space of the system is spanned by the vectors

$$|n_1, n_2\rangle = |n_1\rangle|n_2\rangle = \frac{\left(a_1^\dagger\right)^{n_1} \left(a_2^\dagger\right)^{n_2}}{\sqrt{n_1!}\sqrt{n_2!}}|0,0\rangle, \tag{15}$$

where $|0,0\rangle$ is the vacuum state such that $\hat{a}_1|0,0\rangle = \hat{a}_2|0,0\rangle = 0$ and $n_1$, $n_2$ are nonnegative integers. The energy eigenvalue are

$$E_{n_1 n_2} = (n_1 + n_2 + 1)\omega = (n+1)\omega, \quad n = n_1 + n_2 = 0, 1, 2, \ldots. \tag{16}$$

All the states $|r, n-r\rangle$ with $r = 0, 1, 2, \ldots n$ have the same energy, i.e. the energy eigenvalue $E_n$ is $(n+1)$-fold degenerate. Degeneracy in energy spectra indicate that there are symmetries associated with the system. In this case the symmetry in question is $SU(2)$, that allows for transforming eigenstates with the same energy among themselves. It is worthwhile to work it out in some detail.

Let us construct new operators

$$\hat{J}_0 = \frac{1}{2}\left(\hat{a}_1^\dagger \hat{a}_1 - \hat{a}_2^\dagger \hat{a}_2\right); \quad \hat{J}_+ = \hat{a}_1^\dagger \hat{a}_2; \quad \hat{J}_- = \hat{a}_1 \hat{a}_2^\dagger. \tag{17}$$

These are conserved quantities as one can readily verify

$$i\frac{\partial \hat{J}_k}{\partial t} = \left[\hat{J}_k, \hat{H}\right] = 0, \quad k = 0, \pm \tag{18}$$

The operators so defined in (17) close the familiar $SU(2)$ algebra

$$\left[\hat{J}_+, \hat{J}_-\right] = 2\hat{J}_0, \quad \left[\hat{J}_0, \hat{J}_\pm\right] = \pm \hat{J}_\pm. \tag{19}$$

The expressions (17) are known as Schwinger's bosonic realization of the usual angular momentum operators. One can interpret the action of the operators $\hat{J}_+$ and $\hat{J}_-$ in the following way. The motion of the system on a plane surface is composed of oscillations in two directions, say, $x$ and $y$. The operator $\hat{J}_+$ increases the oscillation amplitude in the $x$-direction and decreases the oscillation in the $y$-direction. It can continue doing so until the state of the particle consists only of oscillation in the $x$-direction. On the other hand, the operator $\hat{J}_-$ just does the opposite; it squeezes the orbit of the particle towards oscillation in the $y$-direction.

The quadratic Casimir of the algebra

$$\hat{J}^2 = \hat{J}_0^2 + \frac{1}{2}\left(\hat{J}_+ \hat{J}_- + \hat{J}_- \hat{J}_+\right) \tag{20}$$

can be expressed as

$$\hat{J}^2 = \frac{1}{4}\left(\hat{a}_1^\dagger \hat{a}_1 + \hat{a}_2^\dagger \hat{a}_2\right)\left(\hat{a}_1^\dagger \hat{a}_1 + \hat{a}_2^\dagger \hat{a}_2 + 2\right). \tag{21}$$

In the basis $\{|n_1, n_2\rangle\}$ both the commuting operators $\hat{J}_0$ and $\hat{J}^2$ are diagonal, that is,

$$\hat{J}_0 |n_1, n_2\rangle = \frac{1}{2}(n_1 - n_2)|n_1, n_2\rangle, \tag{22}$$

$$\hat{J}^2 |n_1, n_2\rangle = \frac{n}{2}\left(\frac{n}{2} + 1\right)|n_1, n_2\rangle. \tag{23}$$

Making correspondence with the usual $SU(2)$ representation, let $n = 2j$, $n_1 = j + m$ and $n_2 = j - m$, where $j = 0, 1/2, 1, 3/2, \ldots$ and $m = -j, -j + 1, \ldots j - 1, j$. One identifies

$$|n_1, n_2\rangle = |j + m\rangle|j - m\rangle \equiv |j, m\rangle, \tag{24}$$

such that

$$\hat{J}_0 |j, m\rangle = m |j, m\rangle, \tag{25}$$

$$\hat{J}^2 |j, m\rangle = j(j + m)|j, m\rangle. \tag{26}$$

The energy eigenvalues become

$$E_j = (2j + 1)\omega. \tag{27}$$

Each $E_j$ is $(2j+1)$-fold degenerate. The set of states $\{|j, m\rangle\}$ for each $j$ forms a basis of the standard representation space of $SU(2)$. Thus we conclude that the dynamics of a 2D isotropic oscillator are governed by the $SU(2)$ symmetry.

## 3. Fixing the Immirizi parameter

It is apparent from the above discussion that both the isotropic oscillator and edges of loop quantum gravity share the same $SU(2)$ symmetry. One is thus naturally tempted to represent a QNM frequency on the horizon as a 2D isotropic oscillator, i.e. $\omega_{QNM} = \omega$. The following correspondence between the QNM and the $SU(2)$ edges of loop quantum gravity can then be observed. The ground state of QNM ($j = 0$) having energy $\hbar\omega_{QNM}$ corresponds to the zero eigenvalue ($j = 0$) of the area operator in loop quantum gravity. It is worth noticing that a double quantum jump of the QNM from $j=1$ to $j=0$ corresponds to the detachment of an edge with $j=1$ from the surface, leaving no puncture ($j=0$) behind. Note that the energy released in the transition is equal to $2\hbar\omega_{QNM}$. This would amount to the release of a fermion-antifermion pair in Corichi's approach [14], each particle carrying energy $\hbar\omega_{QNM}$. It can also be noticed that, in the reverse process, the excitation of QNM from the ground state to the $j=1$ state corresponds to the attachment of a $j=1$ edge to the horizon. We immediately observe that if $SU(2)$ is the relevant gauge group and that if $j_{\min} = 1$ processes dominate, the change in mass of the black hole in this transition should be $\Delta M = 2\hbar\omega_{QNM}$ instead of $\hbar\omega_{QNM}$, as in the case of $SO(3)$ [9]. This implies that equation (6) is to be replaced by

$$\Delta M = 2\hbar\omega_{QNM} = \frac{2\hbar \ln 3}{8\pi M}. \tag{28}$$

Following the same steps, as in section 1, we obtain the modified Immirizi parameter

$$\gamma = \frac{\ln 3}{\pi\sqrt{2}}, \quad \text{for } j_{\min} = 1. \tag{29}$$

This is the main result of the present paper which differs from the value reported in [9] and [14] based on $SO(3)$ and $SU(2)$, respectively, by a factor of 2. Remarkably, this value is reminiscent of the fact that even though the two groups are locally isomorphic, $SU(2)$ is a double-covering map of $SO(3)$, that is, a homomorphism that map two points in $SU(2)$ to one point in $SO(3)$. Notably, our value matches with the one obtained in [16] on the basis of supersymmetric spin networks, but for $j_{\min} = 1$ rather than $j_{\min} = 1/2$.

## 4. Conclusion

We emulated an asymptotic quasinormal mode on the horizon with a 2D oscillator which carries the same symbols and transformation properties as those of the edges in loop quantum gravity. This realization appeared to be meaningful for at least two reasons. On one hand, it made explicit how the conversion of areal quanta to matter quanta, and vice versa, takes place in a consistent way. On the other hand, it led us to work out the correct value of $\gamma$ when $j=1$ processes in the $SU(2)$ framework are taken as dominant.

In [14], it was noted that in order to have fermions in theory, one has to stick to $SU(2)$ as the gauge group. If this is the case, then $j=1/2$ punctures are not forbidden in principle, but they must be suppressed and would be something like "primordial punctures". Therefore, one must look for the exact dynamical mechanism that explains how the dominant contribution to the black hole entropy comes from $j=1$ transitions. The results obtained in [16] by supersymmetric extension of the gauge group is surprising. Does this imply that the underlying theory is in fact suprsymmetry? Again, the real answer will come from a deeper understanding of the dynamic processes.

The realization of the asymptotic QNM frequencies as $SU(2)$ systems is within the framework of the black hole spectroscopy initiated by Bakenstein [18]. It clearly allows for the mass (and hence the area) of a Schwarzschild black hole to have an equally spaced discrete spectrum, each level having $(2j+1)$-fold degeneracy. As $\omega_{QNM}$ is characteristic of the black hole by virtue of (5), no frequency emitted by the black hole can be expected as smaller than $\omega_{QNM}$. Furthermore, the ground state of a QNM with energy $\hbar\omega_{QNM}$ contributes nothing to the area and hence to the entropy of a black hole. But, as in any quantum theory, the ground state is important in a complete description of a quantum black hole; a quantum black hole is built up from the ground state.